\documentclass[]{elsarticle}

\usepackage{lineno,hyperref}
\modulolinenumbers[5]
\usepackage{amsmath,empheq}
\usepackage{amsfonts}
\usepackage{amsthm}
\usepackage{amssymb}
\usepackage{graphicx}
\usepackage{hyperref}
\usepackage{soul}
\usepackage[normalem]{ulem}
\usepackage{textcomp}
\usepackage{commath} 
\usepackage{xcolor}

	\newcommand{\ncd}{\newcommand}
	\ncd{\mrm}    {\mathrm}
	\ncd{\beq} {\begin{equation}}
	\ncd{\eeq} {\end{equation}}
	\ncd{\nn}{\nonumber}

	\def\d{{\rm d}}
	
	\def\iiota{\dot\iota}

	\def\basis[#1]{\frac{\partial}{\partial #1}}
	\def\dt[#1]{\frac{\d}{\d #1}}

	\def\iiota{\dot\iota}

	\usepackage{tikz}
	\usetikzlibrary{decorations.pathreplacing,arrows,positioning,angles,calc,external}
	\tikzset{
    		>=stealth',pil/.style={->,thick,shorten <=2pt,shorten >=2pt,}
		}
 	\usepackage{pgfplots}
 	\pgfplotsset{compat=1.3,every axis/.append style={font=\small,thin,tick style={ultra thin}}}
  	\usepgfplotslibrary{fillbetween}

\begin{document}
\begin{frontmatter}

\title{Exact trajectories of charged particles in the Dirac monopole field}

\author[]{C. S. L\'opez-Monsalvo}
		\ead[cslm]{cslm@azc.uam.mx}
\author[]{A. Rubio-Ponce}
		\ead[arp]{arp@azc.uam.mx}

		\address{Departamento de Ciencias B\'asicas, Universidad Aut\'onoma Metropolitana Azcapotzalco
		Avenida San Pablo Xalpa 180, Azcapotzalco, Reynosa Tamaulipas, 02200 Ciudad de 
		M\'exico, M\'exico}

%
\begin{abstract}

We consider the motion of charged particles in the presence of a Dirac magnetic monopole. We use an extension of Noether's theorem for systems with magnetic forces and integrate explicitly the equations of motion. 

\end{abstract}


\end{frontmatter}

\section{Introduction}

The exploration of the properties of magnetic curves -- the trajectories followed by  charged test particles in the presence of a magnetic field --  is a timely problem in mathematical physics and geometry. In particular, it can be shown that (see Proposition 2.1 in \cite{barros2005gauss}) magnetic curves cannot be the geodesics of any affine connection on the Riemannian manifold they propagate.  In this sense, the natural setting for studying such systems from a variational point of view is not that of Riemannian but Finslerian geometry. Nevertheless, the problem of integrating  the non-geodesic problem  of  magnetic curves can be tackled using standard Riemannian techniques together with  a generalization of Noether's theorem \cite{ikawa2003hamiltonian}. 

The simplest non-trivial static magnetic field is that of a  monopole.  Albeit ``physically prohibited'' according to classical electromagnetism  formulated on a topologically trivial region (e.g. Minkowski spacetime), they can be easily realised as topological defects. That is, while the local dynamics of the electromagnetic field is determined by the differential form of Maxwell's equations, its global properties depend, in a crucial manner, on the topology of the  domain where the fields are defined \cite{preskill1984magnetic}. Moreover,  it turns out that the condition for electric charge quantization in vacuum  is equivalent to a non-trivial topology for spacetime itself (see \cite{dirac1931quantised,dirac1948theory,trautman1977solutions,ryder1980dirac}). 

Interestingly, long before the modern tools of symmetry methods for integrating differential equations, Poincare presented the general properties of the magnetic curves of the magnetic monopole in a rather brief note published in 1896 \cite{poincare1896remarques} (cf. McDonald's notes \cite{mcdonald2015birkeland} for a more detailed explanation), arguing that the trajectories should lie on the surface of a cone whose vertex is located at the monopole. Then, in 1960 Lapidus and Pietenpol extended Poincare's result  and obtained the solution in a coordinate system adapted to the cone on which the motion takes place \cite{lapidus1960classical}.  More recently, Sivardi{\`e}re \cite{sivardiere2000classical} revisited the problem and used a rotating frame to reach the solution. Then, Mayrand \cite{mayrand2014particle} generalised the study of the integrability of the particle dynamics on the cones to higher dimensions. 

The manuscript is structured as follows. In section \ref{sec.mcp} we present the magnetic curves problem from a differential geometric Newtonian point of view. In section \ref{sec.fi}, we recall Sternberg's lemma ``\emph{the electromagnetic field determines the symplectic structure of spacetime}'' and lift the field 2-form to phase space to assemble a non-canonical symplectic structure such that the resulting projected Hamiltonian flow  of  free particles corresponds precisely to the magnetic curves on spacetime. Then, recalling the result by Ikawa \cite{ikawa2003hamiltonian}, we present the first integrals to be obtained in section \ref{sec.mm}. Then, providing a set of initial conditions on the ``equatorial plane'' we obtain explicitly the exact trajectories of charged particles moving in the monopole field. We present a graphic analysis showing emergent features such as ``velocity ripples'' associated with the spring-like trajectory of each individual particle. Thus, albeit the solutions are known, the method itself is a novel combination of differential geometry and magnetic integrable systems exhibiting interesting emergent properties.

\section{The magnetic curves problem}
\label{sec.mcp}

Consider an $n$-dimensional ($n\geq 3$) spacetime manifold $(M,g)$. The path followed by a point particle of mass $m$ is a curve $\gamma:I\subset \mathbb{R}\longrightarrow M$ such that Newton's second law
	\beq
	\label{eq.newton}
	f^\sharp = m a \quad \text{with} \quad f^\sharp = g^\sharp(f),
	\eeq
is satisfied. Here $f\in T^*M$ is a 1-form expressing the external forces responsible for the non-conservation of the particle's momentum, i.e. 
	\beq
	f = \frac{\d}{\d \tau} p \quad \text{where} \quad p=g^\flat(\dot \gamma);
	\eeq
 $a\in TM$ is the acceleration vector field associated with the Levi-Civita connection $\nabla$ given by
	\beq
	a = \frac{D}{\d \tau} \frac{\d}{\d \tau} \gamma = \nabla_{\dot \gamma} \dot \gamma \quad \text{with} \quad \dot \gamma \equiv \frac{\d}{\d \tau} \gamma\in TM
	\eeq
and $\tau$ is an arc-length affine parameter for the curve representing its \emph{proper time}. Derivatives with respect to such parameter are computed by 
	\beq
	\frac{\d}{\d \tau} \equiv \pounds_{\dot \gamma}
 	\eeq
where $\pounds_{\dot \gamma}$ is the Lie derivative along the particle velocity vector field.


We have also introduced the metric isomorphisms $g^\flat:TM\longrightarrow T^*M$ together with $g^\sharp:T^*M\longrightarrow TM$ where $g^\sharp \circ g^\flat = {\rm id}_{T_x M}$ and $g^\flat \circ g^\sharp = {\rm id}_{T_x^* M}$ for every $x \in M$ to establish \eqref{eq.newton} as a relation in $TM$. 
	
Expression \eqref{eq.newton} is a mere statement of the compatibility between the Lie derivative and the Levi-Civita connection (cf. \cite{arnold2021topological}),
	\beq
	\pounds_v v^\flat = \left(\nabla_v v\right)^\flat + \frac{1}{2}\ \d g(v,v),
	\eeq
for the velocity vector field of an affinely parametrised curve $\gamma$. As it is well known, a free particle is  determined by $f=0$ and thus satisfies the geodesic equation $\nabla_{\dot \gamma} \dot \gamma = 0$.

An electromagnetic field $F$ on $(M,g)$ is a solution to Maxwell's equations
	\beq
	\label{eq.maxwell}
	\d F = 0 \quad \text{and} \quad \star \d \star F = j
	\eeq 
where $F$ is a closed 2-form associated with the conservation of the electromagnetic flux while $j$ is a closed $(n-1)$-form expressing the \emph{global} conservation of electric charge, namely
	\beq
	\oint_{\partial \Omega^3} F \overset{!}{=} 0 \quad \text{and} \quad \oint_{\partial \Omega^n} j \overset{!}{=} 0,
	\eeq
where $\star$ is the Hodge dual operator while $\partial \Omega^3$ and $\partial \Omega^n$ denote the \emph{boundaries} of 3 and $n$  dimensional regions of $M$, respectively; and we use $\overset{!}{=}$ to denote an equality that is physically required.


In this setting, the Lorentz force 1-form is postulated to be \cite{hehl2003foundations}
	\beq
	f \overset{!}{=} q\ \iiota_{ \dot \gamma} F
	\eeq
or, equivalently,
	\beq
	f^\sharp \equiv q\ \phi(\dot \gamma) = q\ g^\sharp(\iiota_{\dot \gamma} F)
	\eeq
where $\phi:TM \longrightarrow TM$ is a tangent bundle automorphism which, by construction, satisfies the compatibility condition
	\beq
	\label{eq.compat}
	g\left(\phi(v),u\right) = F(v,u)
	\eeq
for every pair of vector fields $v$ and $u$. Here $\iiota$ denotes the interior product of vector fields and differential forms (we use the same convention as in \cite{kobayashi1963foundations}).
In a local coordinate chart, $\phi$ is nothing but the electromagnetic field with an index raised by the metric, namely $\phi_a^{\ b} = g^{cb} F_{a c}$. 

Therefore, the motion of a particle with mass $m$ and charge $q$ in the presence of the field $F$ is given by
	\beq
	\label{eq.newtoneom}
	\nabla_{\dot \gamma} \dot \gamma = \frac{q}{m}\ \phi(\dot \gamma),
	\eeq
 This is the geometric form of the familiar coordinate expression for the Lorentz force ${f^\sharp}^a=  g^{ab}F_{bc}(q \dot \gamma^c)$.

It follows directly from equations \eqref{eq.compat} and \eqref{eq.newtoneom} that the force is orthogonal to the particle velocity
	\beq
	\label{eq.cond1}
	g\left(\phi(\dot \gamma), \dot \gamma \right) = F(\dot \gamma,\dot \gamma) = 0,
	\eeq
and that the norm of the particle velocity is preserved along its own motion
	\beq
	\label{eq.cond2}
	\frac{\d}{\d \tau} \vert \dot \gamma \vert^2 = \pounds_{\dot \gamma} g(\dot \gamma,\dot \gamma) = 2\  g(\nabla_{\dot \gamma}\dot \gamma,\dot \gamma) = 0,
	\eeq
where we have used the fact that for a metic compatible connection
	\beq
	\pounds_v g(u,w) = g(u,\nabla_v w) + g(\nabla_v u,w).
	\eeq

 \emph{The magnetic curves problem} is that of integrating the system of $n$ second order, non-linear, coupled, ordinary differential equations given by \eqref{eq.newtoneom} for a force depending linearly in the particle velocity satisfying \eqref{eq.cond1} and \eqref{eq.cond2}.  The linear dependence of the Lorentz force in the velocity, makes the integral curves of \eqref{eq.newtoneom} non-reversible under the change $q\rightarrow -q$ or $\tau\rightarrow -\tau$ and is analogous to a particular form of the Zermelo navigation problem studied in the context of geodesic motion on Finsler manifolds \cite{gibbons2009stationary,brody2015solution}

\section{First integrals in the Hamiltonian formulation}
\label{sec.fi}

The equations of motion \eqref{eq.newtoneom} are in general hard to integrate. However, as it is usually the case, a significant simplification arises when there are symmetries present in the formulation of the problem, namely symmetries of $g$ and $\phi$. 
For the geodesic problem, Noether's theorem provides us with the standard tool for obtaining first integrals of the geodesic equation. However, it is not directly applicable to the non-geodesic motion of the magnetic curves problem. Here, we will 
 \emph{lift} the problem to phase space and use Sternberg's construction of a non-canonical  symplectic structure $\omega_F\in\Lambda^2(TM)$ generated by the electromagnetic field $F$ \cite{sternberg1977minimal,guillemin1990symplectic} together with the extension of Noether's theorem developed by Ikawa \cite{ikawa2003hamiltonian,ikawa2007motion,ikawa2010motion} to find first integrals for the \emph{non-geodesic} dynamics  of the \emph{free} particle hamiltonian flow associated with $\omega_F$.  Here,  $\Lambda^2(TM)$ denotes the space of 2-forms on $TM$.

Solving  \eqref{eq.newtoneom} is equivalent to obtaining the integral curves of the  \emph{canonical lift} of $\dot\gamma$ given by its prolongation to the tangent bundle \cite{olver2000applications}
	\beq
  	X_h = \mathsf{Pr}(\dot \gamma) \in T(TM)
	\eeq
satisfying Hamilton's equations
	\beq
	\label{eq.hamilton}
	\iiota_{X_h} \omega_F = \d h.
	\eeq
Here, $\omega_F$ is the 2-form given by
	\beq
	\label{eq.omega}
	\omega_F = \omega - q \pi^*(F), 
	\eeq
where the standard symplectic form $\omega$ is given by 
	\beq
	\label{eq.omega0}
	\omega = -\d \pi^*(p) = -\d \pi^*\left[m g^\flat(\dot \gamma) \right]
	\eeq
where $\pi^*:T^*M \longrightarrow T^*(TM)$ is the induced map associated with the bundle projection $\pi:TM \longrightarrow M$ and $h:TM\longrightarrow \mathbb{R}$ is the \emph{free} particle hamiltonian 
 	\beq
	h = \frac{1}{2}m\ g(\dot \gamma,\dot \gamma).
	\eeq
It follows from \eqref{eq.maxwell} and \eqref{eq.omega0} that $\omega$ is  closed 
and non-degenerate. Indeed, $\omega_F$ is a symplectic form on $TM$. 

The evolution of a function  $\lambda:TM \longrightarrow \mathbb{R}$ along the hamiltonian flow $X_h$ is given by the Poisson bracket $\{\ ,\ \}_F$ associated with $\omega_F$, that is
	\beq
	\label{eq.evoh}
	\frac{\d}{\d \tau} \lambda = \pounds_{X_h} \lambda = \{\lambda, h\}_F \equiv \omega_F(X_\lambda,X_h).
	\eeq
Note that  derivations with respect to the proper time of $\gamma$ are given by the corresponding Lie derivative with respect to the lift of $\dot \gamma$. This is the extension of the correspondence between the hamiltonian and geodesic flows to the case of magnetic curves. 

Let us assume that the set of symmetry generators of the metric and the electromagnetic field 
	\beq
	\label{eq.killing}
	K=\left\{\xi\in TM \vert \pounds_\xi g = 0 \quad \text{and} \quad \pounds_\xi F = 0 \right\}
	\eeq
is non-empty. In such case $\#K=s\leq n(n+1)/2$. Then, for each $\xi \in K$ we have
	\beq
	\pounds_\xi F = \iiota_\xi \d F + \d \iiota_\xi F = 0.
	\eeq
Since $F$ is closed, it follows that, at least locally, there is a function $\psi_\xi:U\subset M \longrightarrow \mathbb{R}$ such that
	\beq
	\label{eq.psi}
	\iiota_\xi F = \d \psi_\xi.
	\eeq
Furthermore, the function 
	\beq
	\label{eq.first}
	\lambda_\xi = \iiota_{\dot \gamma} p_\xi - \psi_\xi \quad \text{where} \quad p_\xi = g^\flat(\xi)
	\eeq
is constant along the free particle Hamiltonian flow \eqref{eq.evoh}, namely
	\beq
	\label{eq.noether}
	\frac{\d}{\d \tau} \lambda_\xi = \frac{\d}{\d \tau} \left(\iiota_{\dot \gamma} p_\xi - \psi_\xi \right) = 0 .
	\eeq
 That is, equation \eqref{eq.first} is a first integral of \eqref{eq.newtoneom} associated with the symmetry generator $\xi$. Additionally, since the norm of $\dot \gamma$ is also a constant of motion, we have a set of $s+1$ first integrals for the system. In this sense, we can consider  equation \eqref{eq.noether}  as an extension of Noether's theorem for the problem of magnetic curves.

 \section{The magnetic curves of the Dirac monopole}
 \label{sec.mm}

 Let us consider $(M,g)$ to be Minkowski spacetime in spherical coordinates $(t,r,\vartheta,\varphi)$ with the section $r=0$ removed, $M=\mathbb{R}^4 - \{r=0\}$, with
 	\beq
	g = - \d t \otimes \d t + \d r \otimes \d r + r^2  g_{S^2}
	\eeq
where 
	\beq
	g_{S^2} = \d \vartheta \otimes \d \vartheta + \sin(\vartheta)^2 \d \varphi \otimes \d \varphi
	\eeq
is the standard metric tensor on the 2-sphare $S^2$.  

 The Dirac monopole electromagnetic field is given by
	\beq
	F = k \sin(\vartheta)\ \d \vartheta \wedge \d \varphi.
	\eeq
It is straightforward to verify that $F$ satisfies the sourceless Maxwell's equations. Indeed
	\beq
	\d F = 0 \quad \text{and} \quad \d \star F = \d \left[\frac{k}{r^2}\ \d r \wedge \d t \right] = 0
	\eeq
are closed 2-forms on $M$. However, integrating $F$ over a space-like 2-surface homeomorphic to $S^2$ enclosing the origin at any given coordinate time $t$  yields 
	\beq
	\label{eq.monopoleflux}
	\oint_{S^2} F = 4 \pi k,
	\eeq
which follows from the non-trivial topological nature of $M$. That is, the non-trivial homology class $[S^2] \in H^2(M)$ is associated with the magnetic charge producing the field $F$, albeit in the domain $M\subset \mathbb{R}^4$ the field itself is indeed sourceless. Moreover,  the `spatial' vector fields measured by a static observer are
	\beq
	\vec{E}_{\rm static} = g^\sharp\left[\iiota_{\frac{\partial}{\partial t}} F\right] = 0
	\eeq
and
	\beq
	\vec{B}_{\rm static} = g^\sharp\left[\iiota_{\frac{\partial}{\partial t}} \star F\right] =- \frac{k}{r^2}\ \hat e_{(r)},
	\eeq
where $\hat e_{(r)}$ is the normalised, outgoing, space-like, radial vector field. 

The field $F$ appears to be regular everywhere except at the poles of the 2-sphere, $\vartheta=0, \pi$ where it becomes singular. However this is merely a coordinate issue which is solved by considering the two overlapping charts  covering the entire surface of the sphere except for the north and south poles, respectively. This path leads to the well known Dirac's charge quantization which we will not use in this work \cite{trautman1977solutions,ryder1980dirac}.

Let us write the velocity of a charged particle in $M$ as
	\beq
	\dot \gamma = \dot t \frac{\partial}{\partial t} + \dot r \frac{\partial}{\partial r} + \dot \vartheta \frac{\partial}{\partial \vartheta} + \dot \varphi \frac{\partial}{\partial \varphi}.
	\eeq
Then, the Lorentz force per charge $q$ is given by
	\beq
	\phi(\dot \gamma) = \frac{k}{r^2} \left[\dot\varphi \sin(\vartheta)\  \frac{\partial}{\partial \vartheta} - \frac{\dot\vartheta}{\sin(\vartheta)}\ \frac{\partial}{\partial \varphi} \right]
	\eeq 
and the equations of motion \eqref{eq.newtoneom} are
	\begin{align}
	\label{eq.eoml01}
	\ddot t 		& = 0,\\
	\ddot r 		& =  r\left[ \sin(\vartheta)^2 \dot \varphi^2 +  \dot \vartheta^2\right],\\
	\ddot \vartheta & = \sin(\vartheta) \cos(\vartheta) \dot \varphi^2 - \frac{2}{r} \dot r \dot \varphi + \frac{q}{m} \frac{k}{r^2} \sin(\vartheta) \dot \varphi, \\
	\label{eq.eoml04}
	\ddot \varphi	& = -2 \cot(\vartheta) \dot \vartheta \dot \varphi -\frac{2}{r} \dot r \dot \varphi - \frac{q}{m} \frac{k}{r^2} \frac{\dot \vartheta}{\sin(\vartheta)}.
	\end{align}
	
The   lift of $\dot \gamma$ to $T(TM)$ is
	\begin{align}
	\label{eq.lift}
	\mathsf{Pr}(\dot \gamma) = 	& \ \dot t \frac{\partial}{\partial t} + \dot r \frac{\partial}{\partial r} + \dot \vartheta \frac{\partial}{\partial \vartheta} + \dot \varphi \frac{\partial}{\partial \varphi}\nonumber\\
								&\ +  \ddot t \frac{\partial}{\partial \dot t} + \ddot r \frac{\partial}{\partial \dot r} + \ddot \vartheta \frac{\partial}{\partial \dot\vartheta} + \ddot \varphi \frac{\partial}{\partial \dot\varphi},
	\end{align}
while the symplectic two form $\omega_F$, equation \eqref{eq.omega}, is given by
	\begin{align}
	{\omega_F} = 			& - m\ \d t \wedge \d \dot t + m\ \d r \wedge \d \dot r \nonumber\\
						& + m r^2\left[\d \vartheta \wedge \d \dot \vartheta + \sin^2(\vartheta)\ \d \varphi \wedge \d \varphi\right]\nonumber\\
						& -2 r\left[ m \dot \vartheta\ \d r \wedge \d \vartheta + m \sin^2(\vartheta)\ \dot \varphi \d r \wedge \d \varphi\right.\nonumber\\
						& \left.\qquad \quad +\sin(\vartheta)\left( m r \dot\varphi \cos(\vartheta) + q k\right)\ \d\vartheta \wedge \d \varphi \right]
	\end{align}
and the free particle hamiltonian is simply
	\beq
	\label{eq.freeham}
	h= \frac{1}{2} m \left[-\dot t^2 + \dot r^2 +r^2 \dot \vartheta^2 + r^2 \sin^2(\vartheta) \dot \varphi^2 \right].
	\eeq
The reader can easily verify that the hamilotnian vector field $X_h$ satisfying Hamilton's equations \eqref{eq.hamilton}  corresponds precisely to substituting  \eqref{eq.eoml01} - \eqref{eq.eoml04} into \eqref{eq.lift}. Indeed, Hamilton's equations of motion \eqref{eq.hamilton} are equivalent to \eqref{eq.newtoneom}, i.e.
	\beq
	\iiota_{X_h} \omega_F = \d h \iff X_h = \mathsf{Pr}(\dot \gamma){\large{\vert}_{\nabla_{\dot \gamma} \dot \gamma = \frac{q}{m}\phi(\dot\gamma)}}.
	\eeq

To integrate  the system \eqref{eq.eoml01} - \eqref{eq.eoml04}, we observe that the elements of $K$ [cf. equation \eqref{eq.killing}] correspond to the symmetry generators of $S^2$ together with that of time translations, that is
	\begin{align}
	\xi_1 & = \frac{\partial}{\partial t},\\
	\xi_2 & = \frac{\partial}{\partial \varphi},\\
	\xi_3 & = \cot(\vartheta) \sin(\varphi) \frac{\partial}{\partial \varphi} - \cos(\varphi) \frac{\partial}{\partial \vartheta},\\
	\xi_4 & = \cot(\vartheta) \cos(\varphi) \frac{\partial}{\partial \varphi} + \sin(\varphi) \frac{\partial}{\partial \vartheta}.
	\end{align}
Then,  integrating \eqref{eq.psi} for each symmetry generator yields
	\begin{align}
	\psi_1 & = c_1 \\
	\psi_2 & = \frac{q}{m} k \cos(\vartheta) + c_2,\\
	\psi_3 & = -\frac{q}{m} k \sin(\vartheta) \sin(\varphi)+c_3,\\
	\psi_4 & = -\frac{q}{m} k \sin(\vartheta) \cos(\varphi)+c_4,
	\end{align}
where the $c_i$'s are integration constants. Finally, the first integrals of the system are [cf.  equation \eqref{eq.first}] 
	\beq
	\label{eq.first1}
	\dot t 	 = \lambda_1,
	\eeq
	\beq
	\label{eq.first2}
	r^2 \dot \varphi \sin^2(\vartheta) - \frac{q}{m} k \cos(\vartheta) = \lambda_2, 
	\eeq
	\begin{align}
	\label{eq.first3}
	r^2 \left[\sin(\vartheta) \cos(\vartheta) \sin(\varphi) \dot \varphi - \cos(\varphi) \dot \theta \right] \nonumber\\
	+ \frac{q}{m} k \sin(\vartheta) \sin(\varphi) = \lambda_3
	\end{align}
	\begin{align}
	\label{eq.first4}
	r^2 \left[\sin(\vartheta) \cos(\vartheta) \cos(\varphi) \dot \varphi + \sin(\varphi) \dot \theta \right] \nonumber\\
	+ \frac{q}{m} k \sin(\vartheta) \cos(\varphi) = \lambda_4,
	\end{align}
together with the norm of the velocity [which is equivalent to the conservation of hamiltonian \eqref{eq.freeham}]
	\beq
	\label{eq.first5}
	-\dot t^2 + \dot r^2 +r^2 \dot \vartheta^2 + r^2 \sin^2(\vartheta) = - \lambda_5^2.
	\eeq
	
A long but straightforward algebraic manipulation allows us to express $\varphi$  as a function of $\vartheta$, namely
	\beq
	\label{eq.eqvarphi}
	\tan(\varphi) = \frac{\zeta_1(\vartheta) \lambda_3 - \zeta_2(\vartheta) \lambda_4 }{\zeta_1(\vartheta) \lambda_4 + \zeta_2(\vartheta) \lambda_3}
	\eeq
where
	\beq
	\zeta_1(\vartheta) = q k - m \lambda_2 \cos(\vartheta)
	\eeq
and
	\begin{align}
	\zeta_2(\vartheta) 	=& \left[\left(\lambda_3^2 + \lambda_4^2\right) m^2 + 2kmq\lambda_2 \cos(\vartheta)\right. \nonumber\\
							 &\left.-k^2 q^2-m^2\left(\lambda_2^2+\lambda_3^2+\lambda_4^2 \right)\cos^2(\vartheta)\right]^{\frac{1}{2}}.
	\end{align}
Thus, the corresponding  angular momenta are given by
	\beq
	L_\varphi = m r^2 \dot \varphi = \frac{k q \cos(\vartheta) - m \lambda_2}{\sin^2(\vartheta)}
	\eeq	
and
	\beq
	\label{eq.vartheta}
	L_\vartheta= m r^2\dot \vartheta = - \frac{\zeta_2(\vartheta)}{ \sin(\vartheta)}.
	\eeq
It is easy to see that these are not conserved [cf. equation \eqref{eq.evoh}, above], that is
	\beq
	\frac{\d}{\d \tau} L_\varphi  = -\frac{qk}{\sin^3(\vartheta)} \left[1 + \cos^2(\vartheta) -\frac{2 m \lambda_2 \cos(\vartheta)}{k q}\right] \dot \vartheta
	\eeq
and
	\beq
	\frac{\d}{\d \tau} L_\vartheta =-\frac{\left[qk - m \lambda_2 \cos(\vartheta) \right] \left[k q\cos(\vartheta) - m \lambda_2\right]}{\sin^2(\vartheta) \ \zeta_2(\vartheta)}  \dot \vartheta.
	\eeq
Indeed, the effect of the magnetic field on the non-radial motion of charged particles  is a torque. This also illustrates that the first integrals \eqref{eq.first}, in particular equations \eqref{eq.first2}--\eqref{eq.first4}, do not correspond to the usual conserved momenta as in the canonical case.


Finally, substituting \eqref{eq.first1}, \eqref{eq.first2} and \eqref{eq.vartheta} into \eqref{eq.first5} yields the radial equation
	\beq
	\label{eq.rdot}
	\dot r^2 = \frac{1}{r^2} \left[\frac{k^2 q^2}{m^2} -\left(\lambda_2^2 + \lambda_3^2 + \lambda_4^2 \right) \right] + \lambda_1^2 - \lambda_5^2.
	\eeq
Therefore, since $\varphi$ is determined solely in terms of $\vartheta$,  the magnetic curves in  a Dirac monopole are completely determined by solving the radial equation \eqref{eq.rdot} and then integrating the angular momentum  \eqref{eq.vartheta} while the coordinate time evolves as the solution of \eqref{eq.first1}.

\subsection{Equatorial initial conditions}

From symmetry considerations any given initial condition can be placed in the equatorial plane. Thus, let us consider the set of initial conditions
	\begin{align}
	\dot r(0) = v_r, \quad 
	r(0) = r_0,\\
	\dot \varphi(0) = \omega_\varphi,\quad 
	\varphi(0) = 0,\\
	\dot \vartheta(0) = 0, \quad
	\vartheta(0) = \pi/2.
	\end{align}
The constants of motion can be expressed in terms of the initial data as
	\begin{align}
	\label{eq.const1}
	\lambda_1 &= \sqrt{1+ v_\varphi^2 + v_r^2},\\
	\label{eq.const2}
	\lambda_2 &= -r_0^2 \omega_\varphi,\\
	\label{eq.const3}
	\lambda_3 &= 0,\\
	\label{eq.const4}
	\lambda_4 &= -qk/m,\\
	\label{eq.const5}
	\lambda_5 &= 1.
 	\end{align}
Here, the last constant represents the speed of light normalisation for a timelike trajectory and  the equatorial initial velocity is defined as
	\beq
	v_\varphi = r_0 \omega_\varphi.
	\eeq

Using \eqref{eq.const1} - \eqref{eq.const5} to integrate \eqref{eq.rdot}  and \eqref{eq.vartheta} we have
	\beq
	\label{eq.solr}
	r^2 = (v_\varphi^2 + v_r^2) \tau^2 - 2 r_0 v_r \tau + r_0^2
	\eeq
and
	\beq
	\label{eq.solvartheta}
	\cos(\vartheta) = -kq\ \frac{ {L}_0}{\ell^2} \left[1-\cos\left(\alpha(\tau) \right)  \right],
	\eeq
respectively. Here the function $\alpha(\tau)$ is given by
	\begin{align}
	\alpha(\tau) =\frac{\ell}{{L}_0} \left[\arctan\left(\frac{[ v_\varphi^2  + v_r^2] \tau + r_0v_r}{r_0 v_\varphi} \right) \right.\nonumber\\
	\left.- \arctan\left(\frac{v_r}{v_\varphi}\right) \right],
	\end{align}
with the angular momentum constants defined as
	\beq
	\ell^2 =k^2 q^2 + {L}_0^2 \quad \text{and} \quad {L}_0 = m r_0  v_\varphi.
	\eeq
Note that equation \eqref{eq.solvartheta} reveals that the constant $k$, quantifying the monopole strength [cf. equation \eqref{eq.monopoleflux}], should have units of action per charge. 

Substituting \eqref{eq.solvartheta} together with the constants \eqref{eq.const2}-\eqref{eq.const4} into \eqref{eq.eqvarphi} we obtain
	\beq
	\label{eq.solvarphi}
	\tan(\varphi) = -\frac{ L_0 \ell\sin\left(\alpha(\tau) \right)}{L_0^2 \cos\left(\alpha(\tau) \right) + k^2 q^2}.
	\eeq
Finally, the time coordinate  is affinely related to proper time as
	\beq
	\label{eq.solt}
	t = \sqrt{1+v_\varphi^2 + v_r^2}\ \tau.
	\eeq

It can be directly verified that, indeed,  \eqref{eq.solr}, \eqref{eq.solvartheta}, \eqref{eq.solvarphi} and \eqref{eq.solt} solve the equations of motion  \eqref{eq.eoml01} - \eqref{eq.eoml04}. Therefore, we have obtained the explicit solution to the magnetic curves problem for the Dirac monopole.

\subsection{Some features of the magnetic curves}

\begin{figure}
\includegraphics[width=1\columnwidth]{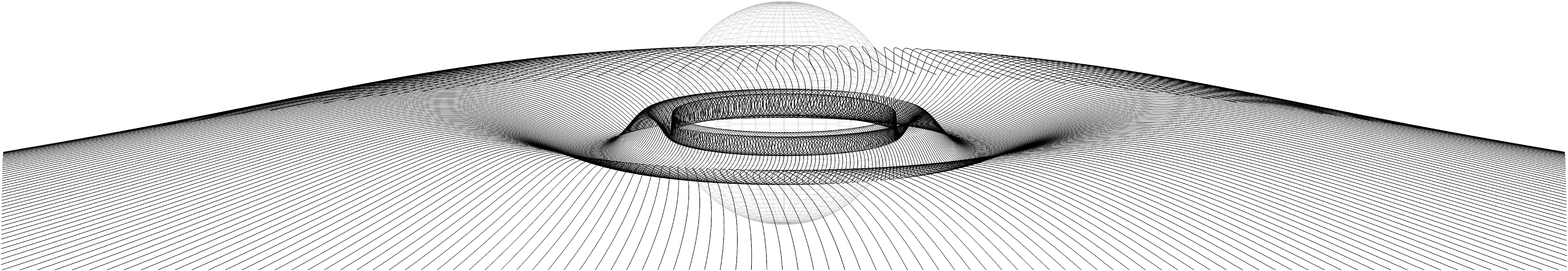}\\
\includegraphics[width=1\columnwidth]{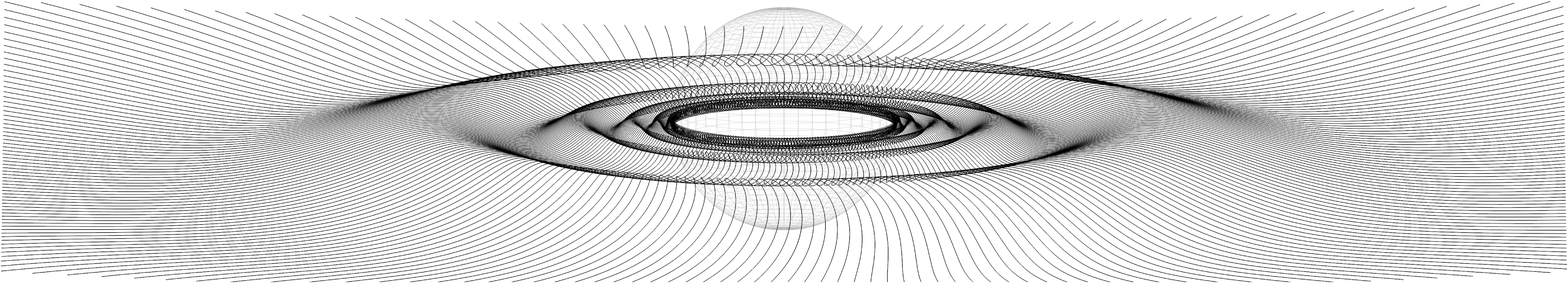}\\
\includegraphics[width=1\columnwidth]{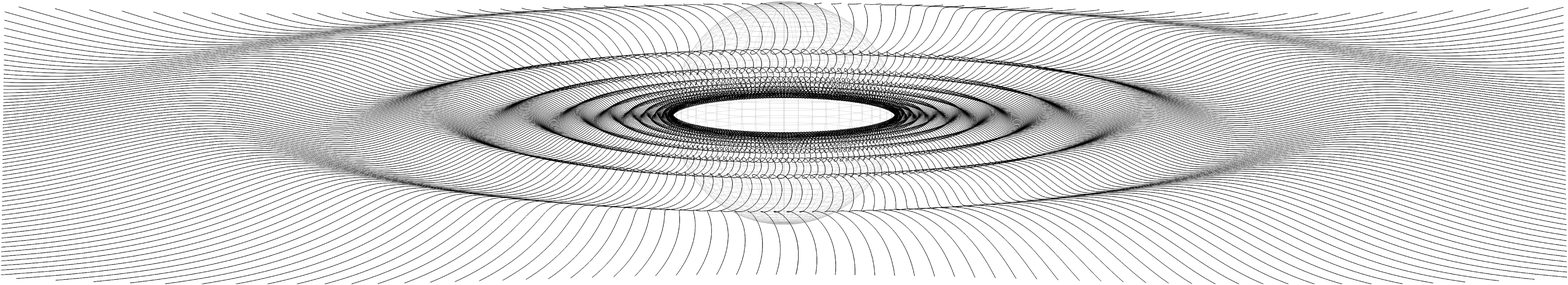}\\
\includegraphics[width=1\columnwidth]{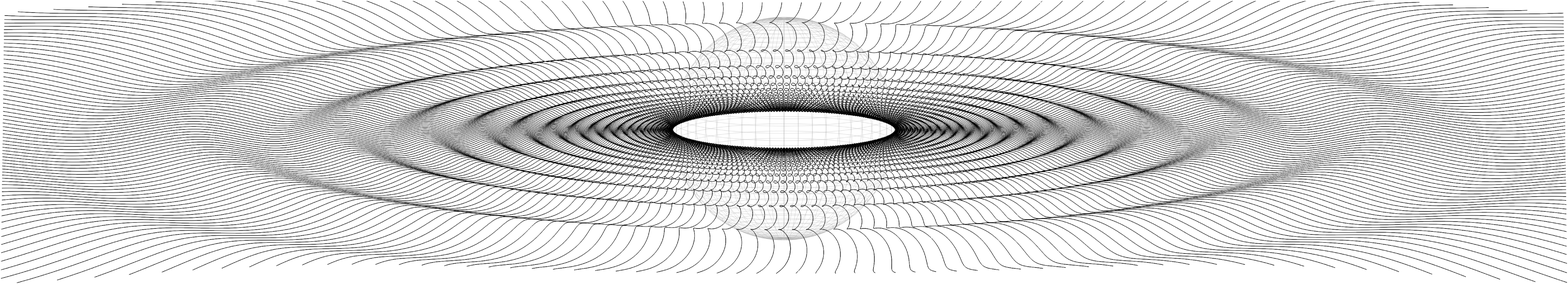}\\
\includegraphics[width=1\columnwidth]{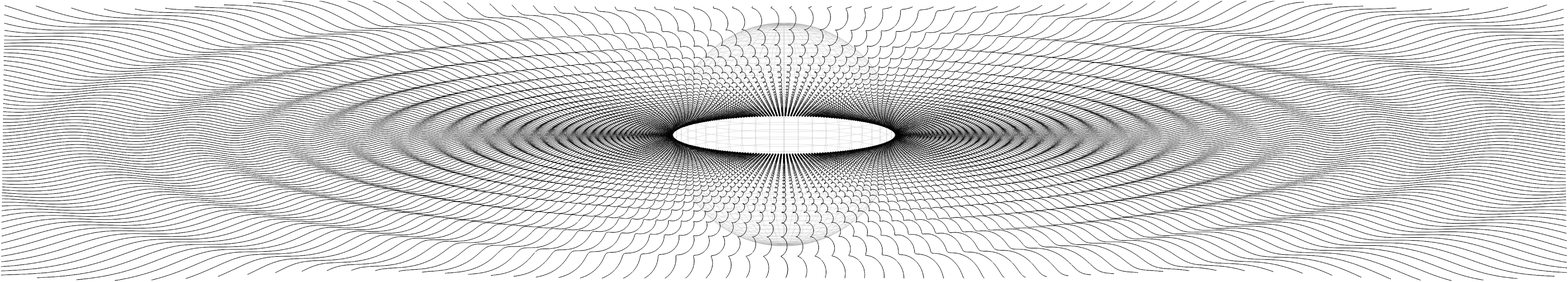}
\caption{The magnetic curves of the Dirac monopole. From top to bottom, we show the curves corresponding to initial velocity $v_\varphi=1/10, 1/25, 1/50, 1/100$ and $1/200$, tangent to the equatorial plane. Here, $r_0=1$. }
\label{fig.curves}
\end{figure}

\begin{figure}
\includegraphics[width=1\columnwidth]{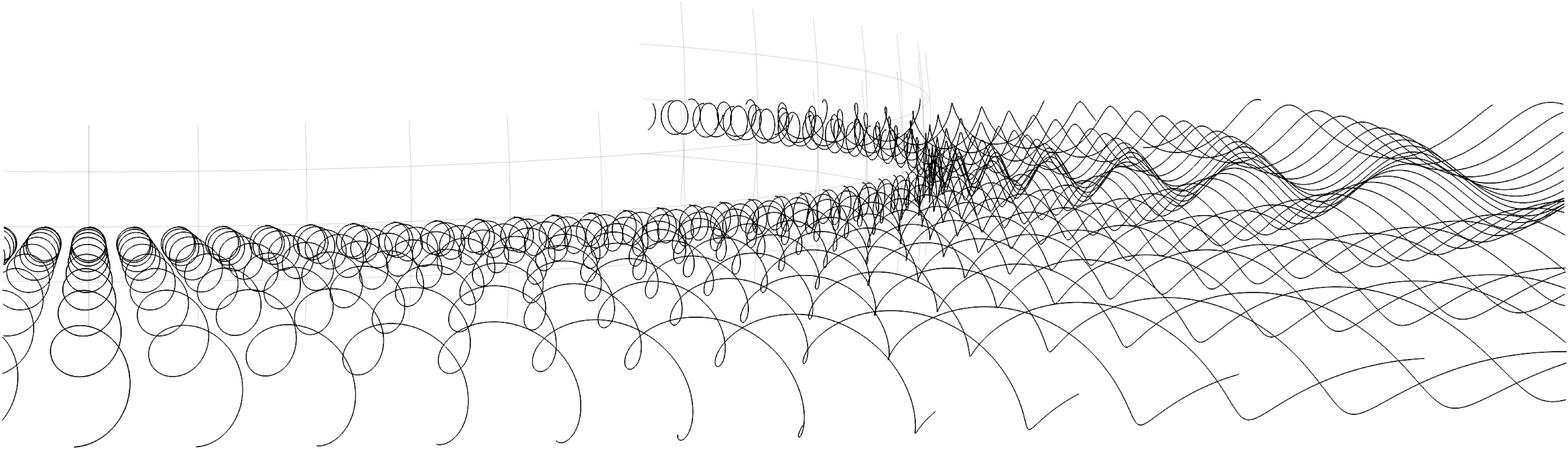}\\
\includegraphics[width=1\columnwidth]{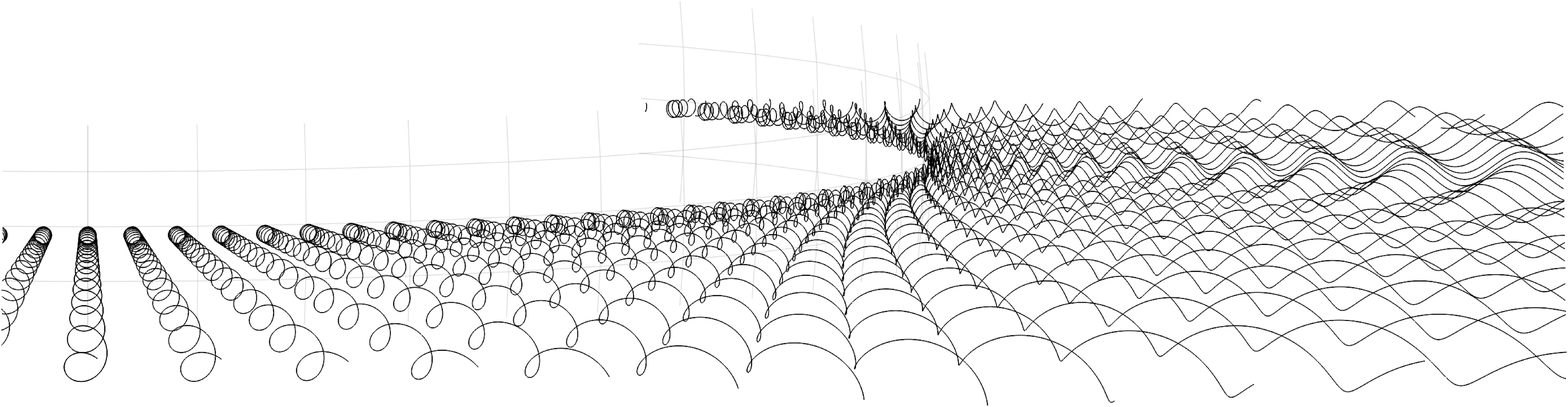}\\
\caption{Detail of the magnetic curves shown in FIG. \ref{fig.curves} corresponding to $v_\varphi = 1/50$ (top) and $v_\varphi=1/100$ (bottom). }
\label{fig.curveszoom}
\end{figure}

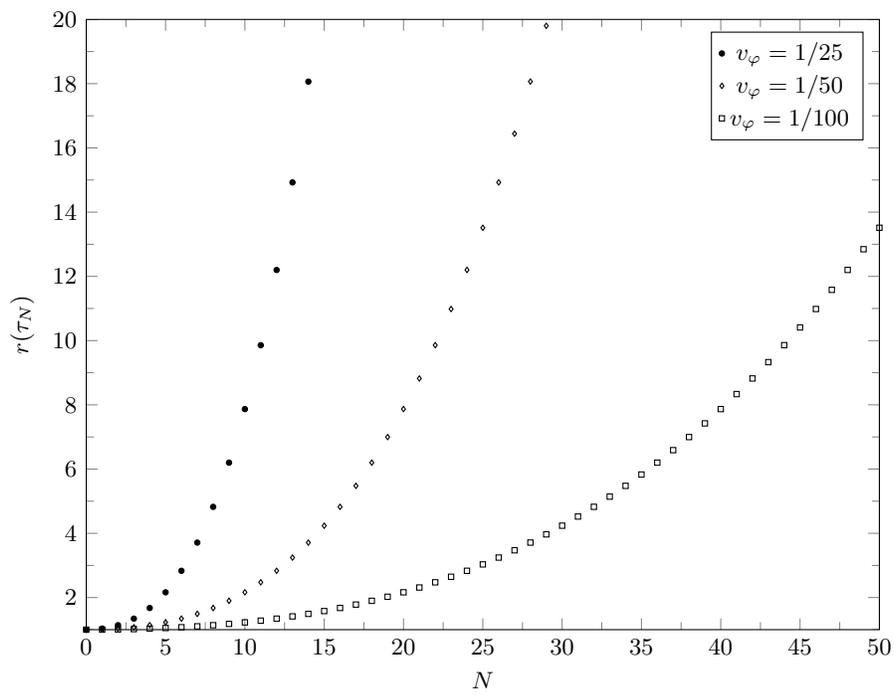
\begin{figure}
        \begin{center}
        \begin{tikzpicture}
        	\begin{axis}[
		xmin=0,xmax=50,
		xlabel={$N$},
		ymin=1,ymax=20,
		ylabel={$r(\tau_N)$},
		minor tick num=1,
		ticklabel style={font=\small},
		scale only axis=false,
		width=1\columnwidth,
        		height=0.8\columnwidth,
		]
		\addplot[only marks, mark = *, color = black, mark size = 1pt] table {zeroes_25.dat};\addlegendentry{$v_\varphi=1/25$};
		\addplot[only marks, mark = diamond, color = black, mark size = 1pt] table {zeroes_50.dat};\addlegendentry{$v_\varphi=1/50$};
		\addplot[only marks, mark = square, color = black, mark size = 1pt] table {zeroes_100.dat};\addlegendentry{$v_\varphi=1/100$};		
		\end{axis}
	\end{tikzpicture}
        \end{center}
        \caption{Radial spreading of the apparent oscillations shown in FIG. \ref{fig.curves}. The dots represent the radial distance for each cycle where $\vartheta(\tau_N) = \pi/2$ for various initial tangential velocities.}
        \label{fig.oscillations}
        \end{figure}

In FIG. \ref{fig.curves} we show the magnetic curves of the Dirac monopole for various initial data. Discarding dissipative effects due to radiation, which is easily justified for non-relatvistic velocities \cite{stratton2007electromagnetic,jackson1999classical} and as noted by Poincare \cite{poincare1896remarques}, the trajectories of charged particles in the monopole field are constrained to lie on a cone with apex at the location of the monopole. This can be directly verified by noting that the  aperture is easily determined to be the constant
	\beq
	\theta_{\rm aperture}^\pm = \frac{\pi}{2} \pm \arctan\left(\frac{ \ell^2 - 2 L_0^2}{2 k q L_0}\right),
	\eeq
where $\pm$ denotes the direction of the initial velocity $v_\varphi$ (cf. FIG. \ref{fig.curveszoom}). This result is in well agreement with that of Zia \cite{zia1979classical}.

Moreover,  there is a minimum for the radial distance reached by a charged particle given by
	\beq
	r_{\rm min}^2 =\left(  \frac{v_\varphi^2}{v_r^2 + v_\varphi^2} \right) r_0^2.
 	\eeq
	
 Finally, note that the emergent wave pattern shown in FIG. \ref{fig.curves} is a consequence of the elongation of the spring-like trajectories and the ever decreasing angular velocities. Indeed,
	\beq
	\lim_{\tau \rightarrow \infty} \dot \vartheta = 0 \quad \text{and} \quad \lim_{\tau \rightarrow \infty} \dot \varphi = 0,
	\eeq
while each cycle, considering a purely tangential initial velocity, occurs every time  that $\vartheta(\tau_N) = \pi/2$ (cf. equations (21) - (23) in \cite{mcdonald2015birkeland}), that is
	\beq
	\label{eq.tauN}
	\tau_N = \frac{r_0}{v_\varphi} \tan\left(2 \pi N \frac{L_0}{\ell} \right).
	\eeq
It is worth recalling that the tangential velocity is constrained by the speed of light. Furthermore, since we are neglecting radiation effects due to the acceleration of the charged particles, we are in the non-relativistic limit, that is $v_\varphi\ll 1$. In such case, we can approximate \eqref{eq.tauN} as
	\begin{align}
	\tau_N \approx  \left[\frac{2 \pi r_0^2 m }{k q} - \left(\frac{m}{k q}\right)^3 \pi r_0^4 v_\varphi^2 \right] N \nonumber\\+ \left[	\frac{8}{3}\left(\frac{\pi m}{k q}\right)^3 r_0^4 v_\varphi^2\right] N^3.
	\end{align}
In FIG. \ref{fig.oscillations} we can see the way in which higher tangential velocities yield a very rapid radial spreading, while in the low velocity limit more ``oscillations'' are observed.

\section{Closing remarks}

In this work, we have obtained in a detailed and explicit manner the solution to the magnetic curves problem in the case of a magnetic monopole. The geometric formulation allows for a direct extension to arbitrary Riemannian manifolds and magnetic fields. It is worth noting that, albeit the literature on the subject has a long history,  a thorough analysis based on the symmetries of the problem was lacking. Here, we fill such a gap. In particular, we present the exact analytical expressions in spherical coordinates for the curves together with a visualization of the velocity field, exhibiting an emerging ripple pattern. Moreover, this geometric perspective can be directly extended to include dissipative effects due to radiation by means of contact geometric techniques, which will be the subject of a forthcoming work.

\section*{Acknowledgements}
CSLM is thankful to Alessandro Bravetti and Francisco Nettel for insightful comments during the preparation of the manuscript. The authors gratefully acknowledge SNI-Conacyt Mexico.

\section*{References}
   

\bibliography{ref_monopole}

\end{document}